\documentclass[sigconf,screen,nonacm]{acmart}
\usepackage{enumitem}

\usepackage{xspace}
\usepackage{color, colortbl}  % For highlighting

\usepackage{adjustbox}
\usepackage{siunitx}
\usepackage{tabularx} % Add this in the preamble
\usepackage{enumitem} % for customizing list environments
\usepackage{amsmath}  % for the arrow symbol (optional)
\usepackage[normalem]{ulem}
\useunder{\uline}{\ul}{}

\newcommand{\ie}{{i.e., }}

\newcommand{\datamorgana}{\texttt{DataMorgana}\xspace}

\usepackage{xcolor}
\definecolor{dgrey}{gray}{0.2}

\definecolor{specialblue}{HTML}{018786}

\definecolor{darkyellow}{HTML}{BA8E23}

\usepackage[english]{babel}  
\addto\extrasenglish{

}

\usepackage[capitalize]{cleveref} %to use \cref{} and \Cref{}

\newcommand{\prompt}[1]{\begin{quote}\tt #1 \end{quote}}

\usepackage[role = ACMBlue, skipempty, separator = {\newline}]{credits}

\usepackage{multirow}
\usepackage{arydshln}

\AtBeginDocument{%
  }

\copyrightyear{2025}
\acmYear{2025}
\setcopyright{cc}
\setcctype{by-sa}
\acmConference[SIGIR LiveRAG]{}{July 17, 2025}{Padua, Italy} \acmBooktitle{SIGIR 2025 LiveRAG Challenge, July 17, 2025, Padua, Italy}
\acmDOI{10.48550/arXiv.2506.14516}

\begin{document}

\title{RMIT--ADM$+$S at the SIGIR 2025 LiveRAG Challenge}
\subtitle{G-RAG: Generation-Retrieval-Augmented Generation}

\author{Kun Ran}
\orcid{0009-0004-6708-703X}
\affiliation{%
  \institution{RMIT University}
  \city{Melbourne}
  \country{Australia}
}
\email{kun.ran@student.rmit.edu.au}

\author{Shuoqi Sun}
\orcid{0009-0000-9329-9731}
\affiliation{%
  \institution{RMIT University}
  \city{Melbourne}
  \country{Australia}
}
\email{shuoqi.sun@student.rmit.edu.au}

\author{Khoi Nguyen Dinh Anh }
\orcid{0009-0004-7668-4290}
\affiliation{%
  \institution{RMIT University}
  \city{Melbourne}
  \country{Australia}
}
\email{s3695517@rmit.edu.vn}

\author{Damiano Spina}
\orcid{0000-0001-9913-433X}
\affiliation{%
  \institution{RMIT University}
  \city{Melbourne}
  \country{Australia}
}
\email{damiano.spina@rmit.edu.au}

\author{Oleg Zendel}
\orcid{0000-0003-1535-0989}
\affiliation{%
  \institution{RMIT University}
  \city{Melbourne}
  \country{Australia}
}
\email{oleg.zendel@rmit.edu.au}

\begin{abstract}
This paper presents the RMIT--ADM+S winning system in the SIGIR 2025 LiveRAG
Challenge. Our Generation-Retrieval-Augmented Generation (G-RAG) approach
generates a hypothetical answer that is used during the retrieval phase,
alongside the original question. G-RAG also incorporates a pointwise large
language model (LLM)-based re-ranking step prior to final answer generation. We
describe the system architecture and the rationale behind our design choices. In
particular, a systematic evaluation using the Grid of Points approach and
$N$-way ANOVA enabled a controlled comparison of multiple configurations,
including query variant generation, question decomposition, rank fusion
strategies, and prompting techniques for answer generation. The submitted system
achieved the highest Borda score based on the aggregation of Coverage,
Relatedness, and Quality scores from manual evaluations, ranking first in the
SIGIR 2025 LiveRAG Challenge.
\end{abstract}


\maketitle

\section{Introduction}
\label{sec:introduction}

Evaluation campaigns such as the SIGIR 2025 LiveRAG
Challenge~\cite{carmel2025sigir} provide a structured and standardized setting
for researchers and practitioners to develop and evaluate different
Retrieval-Augmented Generation (RAG) approaches on a common dataset using shared
metrics. These campaigns enable fair comparisons between systems in a controlled
environment. LiveRAG 2025, the first edition of the challenge, required
participants to develop RAG systems using the \texttt{Falcon3-10B-Instruct}
model~\cite{Falcon3} for final answer generation, thereby standardizing the
generation component across all submissions. This constraint allows for a more
focused evaluation of the retrieval components and prompts, as the generation
model is fixed and does not introduce variability in the results.\footnote{Other
open-weight LLMs, such as the Llama models, were allowed in other components of
the system, but only up to their 10B versions.} 

In this context, the RMIT--ADM+S team submitted \textit{G-RAG}, a system
selected through an internal evaluation process using a Grid of Points (GoP)
approach and $N$-way
ANOVA~\cite{albahemComponentbasedAnalysisDynamic2021}.\footnote{The team
includes members from the RMIT Centre for Human-AI Interaction (CHAI) and the
ARC Centre of Excellence for Automated Decision-Making and Society (ADM+S).}
{The submitted system ranked first in the LiveRAG 2025 Challenge based on the
aggregated manual evaluation scores}, and also achieved the highest individual
scores in the \textit{Coverage} (1.61), \textit{Relatedness} (1.88), and
\textit{Quality} (1.67) metrics. All scores were measured on a 0-2 Likert
scale~\cite{carmel2025sigir}. Our submission builds on previous work and
integrates several components: (1) hypothetical answer generation prior to
retrieval; (2) large language model (LLM)-based query variant
generation~\cite{alaofi2023can,ran2025two}; (3) LLM-based
re-ranking~\cite{sun2025investigation}; and (4) answerability estimation in RAG
systems~\cite{pathiyan2024walert}. We also developed supporting infrastructure
for dynamic cloud-based resource allocation and LLM deployment in the AWS
environment, enabling more efficient resource usage.

The remainder of this paper describes the system architecture and the design
choices made to optimize our RAG system for the LiveRAG Challenge:
Section~\ref{sec:system_architecture} presents the system architecture;
Section~\ref{sec:in-house-evaluation} describes our GoP-ANOVA-based run
selection method; and Section~\ref{sec:conclusion} concludes the work.

The source code for both the system and the evaluation framework are publicly
available at \url{https://github.com/rmit-ir/G-RAG-LiveRAG}.

\section{G-RAG System Architecture}
\label{sec:system_architecture}

\begin{figure*}[t]
      \centering
      \includegraphics[width=0.97\textwidth]{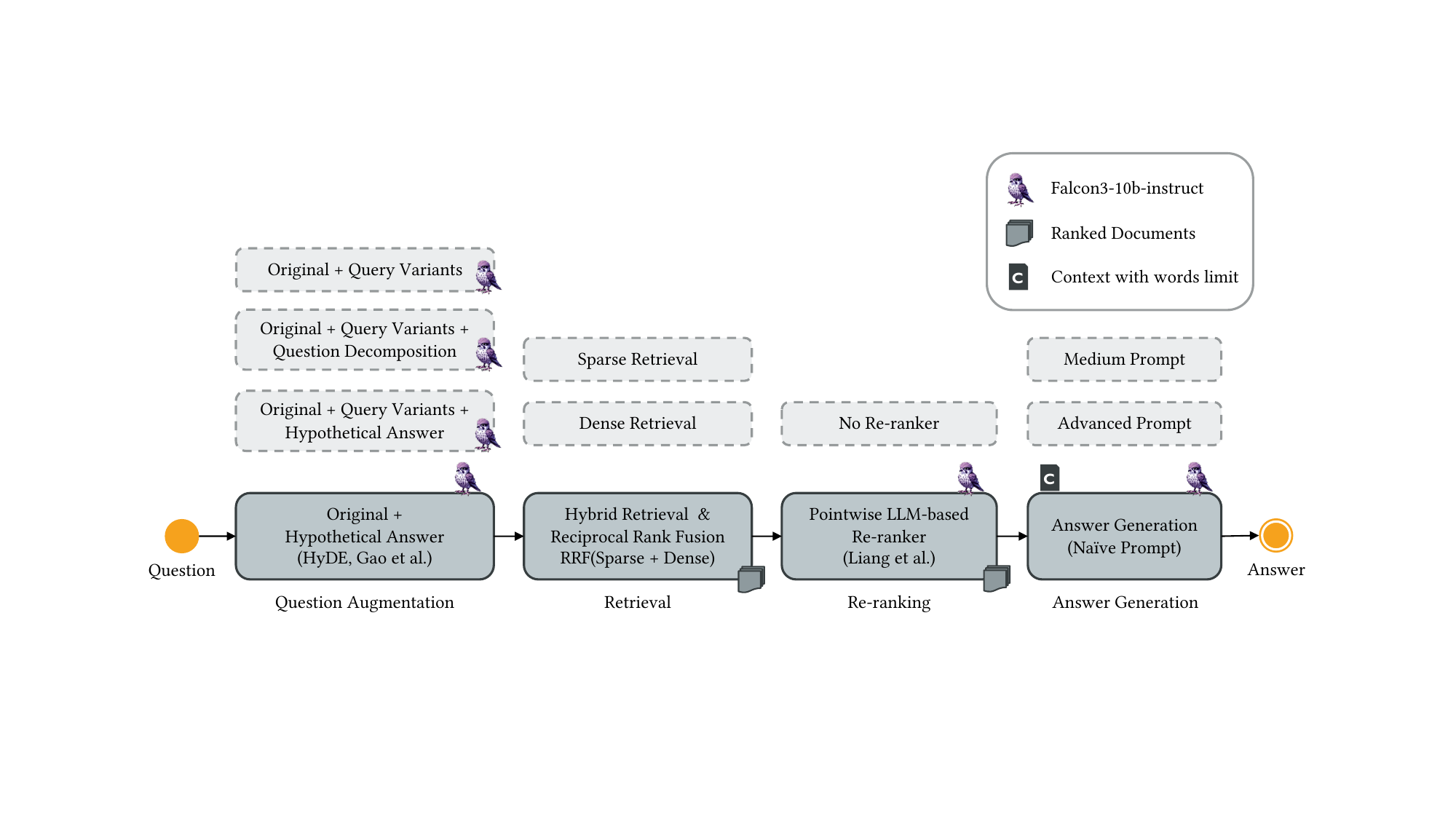}
      \caption{Pipeline of our G-RAG approach. Components used in the final
            selected run are shown with solid borders; components analyzed
            during in-house evaluation are shown with dashed borders.}
      \label{fig:live_rag_pipelines}
\end{figure*}

Figure~\ref{fig:live_rag_pipelines} illustrates the pipeline of our proposed
approach and the system used in our submitted run. In addition to the retrieval
and answer generation stages of a RAG system, G-RAG includes two additional
stages: question augmentation and re-ranking. For all components involving LLMs,
we used the same fixed open-weight model: \texttt{Falcon3-10B-Instruct}. The
prompts used throughout the system are provided in
Appendix~\ref{sec:appendix_prompts}.

\subsection{Question Augmentation}
\label{sec:question_augmentation}

Intuitively, the retrieval phase may benefit from generating queries that offer
complementary ways of retrieving relevant information from the corpus. We
experimented with three query augmentation approaches: query variants, question
decomposition, and hypothetical answer generation.

\emph{Query Variants.} This approach generates multiple query variations from
the original question to be used as search queries. We adapted prompts from
prior work~\cite{ran2025two,alaofi2023can} to better suit the
\texttt{Falcon3-10B-Instruct} model.

\emph{Question Decomposition.}
During manual examination and evaluation of the dataset -- particularly when
identifying \emph{tricky questions} (see
Section~\ref{subsubsec:dataset_preparation}) -- we observed cases where query
variants were insufficient for handling complex questions that involve
comparisons or require multiple aspects. To address this, we designed a process
involving multiple calls to the LLM to decompose such questions into
sub-queries. We begin by identifying the essential components of the question
using shorthand entity annotation. These components are then used to rephrase
the original question into a more detailed and human-readable form. Finally, we
developed a lightweight classifier based on the rephrased question to decide
whether to apply query variants or to reformulate sub-questions as independent
queries. Since the model is relatively small, each step is efficient; the
classifier runs in approximately 100ms on an NVIDIA L40S GPU.

\emph{Hypothetical Answer Generation.}
Inspired by the Hypothetical Document Embeddings (HyDE) approach proposed by
\citet{gao2023precise}, we use the Falcon LLM to generate a ``hypothetical
answer.'' Since the initial prompt did not yield an answer for certain
questions, it was modified to encourage the model to be more flexible and
produce a response that ``could be true.'' The hypothetical answer is then
treated as a search query: it is added to the list of queries and used in the
subsequent retrieval stage.

\subsection{Retrieval}
\label{subsec:retrieval}

We used the retrieval services provided by the organizers. Documents were split
into sentence-based chunks with a maximum length of 512 tokens using the
LlamaIndex sentence splitter. For each query, we applied both sparse and dense
retrieval methods. The results were then merged into a single ranked list using
Reciprocal Rank Fusion (RRF)~\cite{cormack2009reciprocal} with $k=60$ (hereafter
referred to as hybrid retrieval). Documents retrieved across all queries were
combined into a unified list for the subsequent re-ranking stage. For sparse
retrieval, we sent each search query directly to the provided OpenSearch
service, which applies Okapi BM25~\cite{robertson1999okapi} over a prebuilt
inverted index. For dense retrieval, we used the provided Pinecone service to
search over sentence chunks embedded as 512-dimensional vectors using E5-base
embeddings~\cite{wang2024textembeddingsweaklysupervisedcontrastive}. The search
query was encoded using the same embedding model (\texttt{intfloat/e5-base-v2})
and sent to the Pinecone service to retrieve documents ranked by cosine
similarity.

\subsection{Re-ranking}

We employ a Pointwise LLM-based re-ranker, leveraging the \emph{likelihood} of
the model to generate ``Yes'' -- extracted from the associated token logits --
as an indicator of document relevance to the query, following the method
proposed by \citet[p.~21]{liang2023holistic}. For each document, we prompt the
LLM to determine whether it contains the necessary information to answer the
question. Documents are then ranked by the resulting \emph{likelihood} scores,
and those scoring below a threshold of $0.5$ are discarded.

\subsection{Answer Generation}
\label{sec:answer_generation}

To generate the final answer, we describe the task in system prompt and include
the retrieved documents along with the original question in the user prompt. We
created three sets of system and user prompts and selected the best-performing
one based on the \emph{champion} configuration (see
Section~\ref{subsec:grid-of-points}).

\emph{Na\"ive Prompt.} We used the similar system prompt used by ChatGPT
``\texttt{You are a helpful assistant},'' combined with a simple task
instruction ``\texttt{Answer the question based on the provided documents}.''

\emph{Medium Prompt.} In addition to minor refinements, this version explicitly
instructed the model to respond with ``I don't know'' when the answer was not
present in the context, following suggestions from prior
work~\cite{pathiyan2024walert,filice2025generating}. This instruction aligns
with the LiveRAG Relevance evaluation metric, where a response of ``I don't
know'' receives a score of $0$, while incorrect answers receive a score of $-1$.
It may also contribute to improved faithfulness by discouraging unsupported
generations.\footnote{Evaluation details are available at
\url{https://liverag.tii.ae/challenge-details.php}.}

\emph{Advanced Prompt.} This version incorporated prompting techniques from
\citet{tamber2025benchmarking} to further discourage hallucinated content and
emphasize faithful generation.

\subsection{Submitted Run Configuration}
The submitted run used hypothetical answer generation for query augmentation (G),
hybrid retrieval, LLM-based re-ranking (R), and the na\"ive prompt for answer
generation (AG). This configuration corresponds to the main horizontal pipeline shown
in Figure~\ref{fig:live_rag_pipelines}.

\section{In-House Evaluation}
\label{sec:in-house-evaluation}

This section presents our systematic in-house evaluation. We describe the setup,
including the test dataset, the RAG system interface, and the LLM-based
evaluation method, as well as the methodologies used for system optimization and
live challenge submission (\ie Grid of Points and ANOVA).

\subsection{Experimental Setup}

\subsubsection{Dataset Preparation}
\label{subsubsec:dataset_preparation}

\begin{table}[tp]
      \caption{Configuration used to generate the synthetic test collection with
            \datamorgana.}
      \label{tab:dm_config}
      \centering
      %\small \tabcolsep=0.06cm
      \resizebox{0.99\columnwidth}{!}{
            
\begin{tabular}{cccS}
        \toprule
                                                  &  Category Name    &  Value & { Probability} \\
        \midrule

        \em \multirow{2}{*}{User Categories}      &
        \multirow{2}{*}{Expertise}                & Expert
                                                  & 0.5                                                               \\
                                                  &
                                                  & Novice                & 0.5
        \\
        \midrule
        \em \multirow{14}{*}{Question Categories} &
        \multirow{2}{*}{Factuality}               & Factoid
                                                  & 0.5                                                               \\
                                                  &
                                                  & Open-ended            & 0.5                                       \\
        \cmidrule{3-4}
                                                  &
        \multirow{2}{*}{Premise}
                                                  & Direct
                                                  & 0.5
        \\
                                                  &
                                                  & With-premise          & 0.5
        \\
        \cmidrule{3-4}
                                                  &
        \multirow{4}{*}{Phrasing}
                                                  &
        concise-and-natural                       & 0.25
        \\
                                                  &
                                                  & verbose-and-natural   & 0.25
        \\
                                                  &
                                                  & short-search-query    & 0.25
        \\
                                                  &
                                                  & long-search-query     & 0.25
        \\
        \cmidrule{3-4}
                                                  &
        \multirow{2}{*}{Linguistic variation}
                                                  & similar-to-document   & 0.5
        \\

                                                  &
                                                  & distant-from-document & 0.5
        \\
        \cmidrule{3-4}
                                                  &
        \multirow{4}{*}{multi-doc}
                                                  & single-doc
                                                  & 0.4                                                               \\
                                                  &
                                                  & comparison            & 0.2                                       \\
                                                  &
                                                  & multi-aspect          & 0.3
        \\
                                                  &
                                                  & three-doc             & 0.1                                       \\
        \bottomrule
\end{tabular}

      }
      \vspace*{-1em}
\end{table}

We utilized \datamorgana to construct our dataset of question--answer pairs.
Building on the setup presented in the original paper
\cite{filice2025generating}, we created a configuration file
(Table~\ref{tab:dm_config}) comprising one user category and five distinct
question categories. Our configuration remains largely consistent with that
detailed by \citet{filice2025generating}, with a minor adjustment to the
\emph{multi-doc} field. This field is intended to regulate the number of
documents used for question generation, particularly relevant when exploring
whether questions can be formulated using information drawn from multiple
documents.

Multiple datasets were generated at varying sizes: $2$, $5$, $50$, $100$, $200$,
$500$, and $1{,}000$ questions. The smaller datasets were used during the
initial development and debugging phases of the RAG system, while the larger
datasets supported a more extensive evaluation of the system and its individual
components. The largest dataset ($1{,}000$ questions) was specifically used for
stress testing and benchmarking runtime efficiency to ensure the system could
operate within the time constraints of the live event.

To support the development and evaluation of our RAG system, we curated several
smaller datasets with varying levels of complexity. We first constructed a
baseline RAG system that used the original question as the query, combined with
sparse retrieval and an initial answer generation prompt. This system was
applied to a set of 500 questions to obtain initial evaluation results. We then
analyzed the outputs to identify \emph{tricky questions}—those that the baseline
system answered incorrectly or incompletely. Questions receiving the lowest
Relevance and Faithfulness scores were further marked as \emph{challenging
questions}. In total, we identified 179 \emph{tricky questions} and 15
\emph{challenging questions}. To support the GoP and ANOVA analyses described in
Sections~\ref{subsec:grid-of-points} and~\ref{subsec:anova-analysis}, we
randomly selected 15 questions from the \emph{tricky questions} and combined
them with 85 questions drawn from a separate dataset to construct the main test
set.

These datasets were employed iteratively throughout the development process to
examine the behavior of the RAG pipeline and its individual components. After
each evaluation cycle, we reviewed the results to detect recurring failure
patterns and refine the system accordingly. For example, in a question ``total
funding amount digital health startups ryse,'' the term ``ryse'' refers to a
startup. However, the query augmentation module frequently miscorrected it to
``rise,'' misinterpreting it as a typographical error. This illustrates a
negative optimization effect caused by query rewriting. Another example is the
question ``How does the artwork 'For Proctor Silex' create an interesting visual
illusion for viewers as they approach it?'' Here, the named entity ``For Proctor
Silex'' is rare and difficult to retrieve from the corpus. As a result, the
relevant document was ranked low, leading to an inaccurate answer. These cases
highlight the need for retrieval components to return a sufficiently diverse set
of documents, especially for queries containing rare or ambiguous named
entities.

\subsubsection{Combined RAG System}
\label{subsubsec:system-interface}

After testing individual components across multiple RAG system variants, we
developed a modular, configurable RAG system that supports the specification of
different components and settings for each stage of the pipeline. This combined
system enables controlled experimentation by allowing individual components to
be modified independently, while holding others constant. As such, it serves as
a foundational tool for systematic evaluation and parameter exploration.

The configurable aspects of the combined RAG system include:

\begin{itemize}[itemsep=2pt, topsep=4pt, parsep=0pt, partopsep=0pt, left=1em]
      \item Answer Generation Prompts: Introduced in
            Section~\ref{sec:answer_generation}.
      \item Question Augmentation: As detailed in Section
            \ref{sec:question_augmentation}.
      \item Query Variants Generation Prompts: Similar in structure to the
            answer generation prompts; these are used to generate query variants
            within the \emph{Question Augmentation} module.
      \item Number of Query Variants: Controls how many variants are generated
            when employing query augmentation.
      \item Retrieval: As described in Section~\ref{subsec:retrieval}.
      \item Re-ranker: Optionally applies a pointwise LLM-based re-ranker.
            Pointwise re-ranking has shown competitive
            performance~\cite{sun2025investigation}.
      \item Number of Documents Retrieved: Specifies how many documents are
            retrieved for each query prior to re-ranking.
      \item Context Words Length Limits: Retrieved documents are concatenated
            and truncated to a specified number of words before being passed to
            the LLM for answer generation.
\end{itemize}

\subsubsection{LLM-based Evaluation}
\label{subsubsec:llm-evaluation}

This evaluation method follows the guidelines outlined on the LiveRAG Challenge
Details website.\footnote{\url{https://liverag.tii.ae/challenge-details.php}} To
approximate the official evaluation approach, we used Claude 3.5
Sonnet,\footnote{\url{https://www.anthropic.com/news/claude-3-5-sonnet}}
accessed via Amazon Bedrock,\footnote{\url{https://aws.amazon.com/bedrock/}} to
evaluate our system outputs. While the exact prompts and procedures used by the
LiveRAG organizers are not publicly disclosed, we aimed to replicate the setup
as closely as possible. Our evaluation prompt instructed the LLM to compare the
generated answer against the reference answer provided by \datamorgana. To
prevent interference between metrics, we used separate prompts and independent
API calls to obtain relevance and faithfulness scores.

This evaluation procedure was designed to assess both overall performance (i.e.,
average scores across all questions) and per-question performance. These scores
were further used to compare the relative effectiveness of different RAG system
variants.

\subsection{Grid of Points}
\label{subsec:grid-of-points}

\begin{table*}[tb]
      \caption{$N$-way ANOVA results. This analysis utilized the average-level
            LLM-based evaluation method described in
            Section~\ref{subsubsec:llm-evaluation}. A component has a
            statistically significant impact if its probability value,
            PR(\textgreater{}F), is less than 0.05. The header lines are as
            follows: Component: The component being analyzed, refer to
            Section~\ref{subsubsec:system-interface}. SS: Sum of Squares. MS:
            Mean Square. DF: Degrees of Freedom. F: F-statistic.
            PR(\textgreater{}F): Probability value greater than F, \ie
            $p$-value. $\omega_p^2$: Partial Omega Squared indicating effect
            size.}
      \label{tab:n-way-anova}
      \resizebox{.99\textwidth}{!}{
            \begin{tabular}{@{}lSSSSSS}
    \toprule
    Component                                                                       & {SS}   & {DF} & {MS}   & {F}      & {PR(\textgreater{}F)} & {$\omega_p^2$} \\
    \midrule
    Question Augmentation                                                           & 0.3193 & 2    & 0.1596 & 125.6920 & < 0.0001*             & 0.7571         \\
    Query Variants Generation Prompts : Retrieval                                   & 0.0062 & 1    & 0.0062 & 4.9099   & 0.0297*               & 0.0472         \\
    Query Variants Generation Prompts                                               & 0.0030 & 1    & 0.0030 & 2.3892   & 0.1263                & 0.0173         \\
    Context Words Length Limits                                                     & 0.0025 & 1    & 0.0025 & 2.0052   & 0.1608                & 0.0126         \\
    Retrieval                                                                       & 0.0024 & 1    & 0.0024 & 1.8631   & 0.1762                & 0.0108         \\
    Number of Documents Retrieved : Retrieval                                       & 0.0020 & 1    & 0.0020 & 1.6100   & 0.2083                & 0.0077         \\
    Answer Generation Prompts                                                       & 0.0014 & 1    & 0.0014 & 1.1061   & 0.2962                & 0.0013         \\
    Context Words Length Limits : Question Augmentation                             & 0.0008 & 2    & 0.0004 & 0.3261   & 0.7227                & -0.0171        \\
    Context Words Length Limits : Answer Generation Prompts : Question Augmentation & 0.0004 & 2    & 0.0002 & 0.1566   & 0.8553                & -0.0215        \\
    Answer Generation Prompts : Question Augmentation                               & 0.0004 & 2    & 0.0002 & 0.1566   & 0.8554                & -0.0215        \\
    Query Variants Generation Prompts : Number of Documents Retrieved : Retrieval   & 0.0001 & 1    & 0.0001 & 0.1143   & 0.7363                & -0.0113        \\
    Context Words Length Limits : Answer Generation Prompts                         & 0.0001 & 1    & 0.0001 & 0.0491   & 0.8253                & -0.0122        \\
    Query Variants Generation : Number of Documents Retrieved                       & 0.0000 & 1    & 0.0000 & 0.0226   & 0.8808                & -0.0125        \\
    Number of Documents Retrieved                                                   & 0.0000 & 1    & 0.0000 & 0.0118   & 0.9137                & -0.0127        \\
    Residual                                                                        & 0.0978 & 77   & 0.0013 & {--}     & {--}                  & {--}           \\
    \bottomrule
\end{tabular}

      }
      \vspace*{-1em}
\end{table*}

As shown in Figure~\ref{fig:live_rag_pipelines}, our development process
involved testing various parameter combinations, revealing a large configuration
space for each RAG system component. Given the time constraints of the LiveRAG
challenge, identifying an effective configuration was essential. To address
this, we adopted the Grid of Points (GoP)~\cite{Nicola2016LinearMixedModels} and
Analysis of Variance (ANOVA)~\cite{maxwell2017designing, Rutherford2011ANOVA}.
This section describes how we used GoP to identify a \emph{champion}
configuration—defined as the best-performing parameter combination on our
development set. The ANOVA analysis is detailed separately in
Section~\ref{subsec:anova-analysis}.

GoP systematically explores all combinations of parameters within a predefined
configuration space. This process was facilitated by our modular RAG system
described in Section~\ref{subsubsec:system-interface}. We evaluated a total of
96 configurations,\footnote{The 96 configurations result from the Cartesian
product of the options available for each component.} limiting the \emph{Number
of Query Variants} to eight and consistently applying a Pointwise
\emph{Re-ranker} to manage computational constraints. Each configuration was
evaluated using the LLM-based evaluation method, and we ranked the
configurations by their average relevance score. The configuration with the
highest average relevance score was selected as the \emph{champion}. This
configuration achieved an average relevance score of 1.75 and an average
faithfulness score of 0.59, making it one of the top candidates for live
submission. Details of the \emph{champion} configuration are provided in
Section~\ref{subsec:hyde}.

\subsection{$N$-Way ANOVA Analysis}
\label{subsec:anova-analysis}

While the \emph{champion} configuration provided a reliable option for
participating in the live event, our goal remained to further improve system
performance. To this end, we conducted an $N$-way ANOVA analysis, which allowed
us to systematically assess the relative impact of different components and
their interactions. This method guided us in identifying which components have
the most influence on system performance and should therefore be prioritized for
further experimentation.

The ANOVA analysis was performed on the same 96 parameter configurations used in
the GoP evaluation. Since the \emph{Number of Query Variants} and
\emph{Re-ranker} settings were held constant (fixed at eight and Pointwise,
respectively), they were excluded from the ANOVA model and are not shown in
Table~\ref{tab:n-way-anova}. The results presented in
Table~\ref{tab:n-way-anova} indicate that \emph{Question Augmentation} is the
most influential component affecting system performance. The dominant influence
of \emph{Question Augmentation} is evident not only in its statistical
significance ($p$-value$<0.0001$), but also in its effect size. The Partial
Omega Squared ($\omega_p^2=0.7571$) underscores its practical importance,
accounting for the majority of explained variance. In contrast, the negative
$\omega_p^2$ for some components, which occur when an F-statistic is less than
$1.0$, signify the absence of effect~\citep{kroes2023demystifying}. Furthermore,
we observed a statistically significant interaction effect between \emph{Query
Variants Generation Prompts} and \emph{Retrieval}. While neither of these
components had a significant effect on their own, their interaction suggests
that specific combinations of prompts and retrieval methods can jointly
influence performance.

These findings suggest that further improvements beyond the current
\emph{champion} configuration may be achieved by focusing on the most impactful
components. Guided by this insight, we explored an alternative \emph{Question
Augmentation} technique and compared its performance to the original
\emph{champion} configuration.

\subsection{Post-ANOVA Improvement: Hypothetical Answer Generation}
\label{subsec:hyde}

Although \emph{Question Augmentation} emerged as the most impactful component in
the ANOVA analysis, the configuration that omitted augmentation altogether (\ie
the ``None'' parameter setting) yielded the highest Relevance scores.

To further explore improvements in this component, we adopted a hypothetical
answer generation technique inspired by HyDE~\cite{gao2023precise}. This method
introduces \emph{G-RAG}, a variant of our RAG system that employs hypothetical
answer generation as the \emph{Question Augmentation} strategy, replacing the
``None'' setting used in the \emph{champion} configuration. Due to time
constraints, we did not re-run the full GoP or ANOVA tests for this single
parameter change. Instead, we conducted a focused comparison between the
original \emph{champion} configuration and G-RAG. This ad-hoc evaluation was
performed using a sample of 100 questions and measured performance in terms of
both Relevance and Faithfulness.

\begin{table}[tp]
      \centering
      \caption{Comparison of the \emph{champion} configuration and G-RAG across
            100 sampled questions. The table reports how often each system
            achieved higher or tied \emph{Relevance} and \emph{Faithfulness}
            scores on a per-question basis.
            \label{tab:champion-hyde-comparison}}
            \vspace*{-0.3em}
            \resizebox{0.92\columnwidth}{!}{%
            \begin{tabular}{@{}lcc@{}}
    \toprule
    RAG System                             & Relevance Score &
    Faithfulness Score
    \\
    \midrule
    G-RAG (\emph{champion} + HyDE) & \textbf{8}      & 12
    \\
    Champion (GoP)                       & 7               &
    \textbf{14}
    \\
    Ties (Equal Scores)                    & 85              & 74
    \\
    \bottomrule
\end{tabular}
      }
      \vspace*{-1.2em}
\end{table}

We first compared the two systems at the aggregate level using the average
Relevance score. Both the original \emph{champion} and the G-RAG variant
achieved the same average score of 1.75. To gain deeper insights, we conducted a
single-question analysis. As shown in Table~\ref{tab:champion-hyde-comparison},
the G-RAG configuration outperformed the original on 8 questions and
underperformed on 7 in terms of Relevance. For Faithfulness, it scored higher on
12 questions and lower on 14. Given the time constraints and the relatively
balanced performance across metrics, we prioritized Relevance as the primary
evaluation criterion. Based on this, we selected the G-RAG variant as an
improvement over the original \emph{champion} system.

Although HyDE was originally introduced for dense
retrieval~\cite{gao2023precise}, our preliminary experiments on a 50-question
subset indicated that applying hypothetical answers to both sparse and dense
retrieval produced better results than using them with dense retrieval alone. A
more comprehensive investigation of HyDE's integration within RAG pipelines is
left for future work.

In addition to performance considerations, G-RAG also offered improved
efficiency. The ANOVA results indicated that \emph{Context Length Limitation}
had no significant impact on evaluation metrics. However, longer contexts
substantially increased LLM inference time. To reduce latency during the live
event, we limited the input context length to 10k tokens, even though the
original \emph{champion} achieved its best performance with 15k tokens. This
decision was further supported by the \emph{runner-up} configuration, which
demonstrated comparable performance with lower computational cost.

\section{Conclusion and Future Work}
\label{sec:conclusion}

Through a systematic evaluation of our internal runs -- using\linebreak
\datamorgana to generate synthetic datasets and applying a combination of GoP
and $n$-way ANOVA -- we efficiently identified the most cost-effective
combination of the component parameters to maximize system performance within
the time constraints of the LiveRAG 2025 Challenge. This combined approach also
enabled us to prioritize the most impactful components and methodically test and
refine them, resulting in a well-optimized final configuration for our RAG
system. Based on this evaluation, we selected G-RAG as our submitted run. The
system achieved a Relevance score of 1.199 and a Faithfulness score of 0.477,
ranking third on the private LiveRAG 2025 leaderboard and selected as one of the
top four finalists based on the final manual evaluation \cite{carmel2025sigir}. \emph{G-RAG was
ultimately announced as the winning system, validating the effectiveness of our
design and evaluation strategy.}

Future work will focus on a more detailed analysis of system components and
specific cases, including unanswerable or out-of-knowledge-base
questions~\cite{pathiyan2024walert}. We also plan to integrate post-retrieval
Query Performance Prediction (QPP) to dynamically identify which questions would
benefit from query variant expansion~\cite{zendel2021is, ran2025two}.

\section{Ethical Considerations}

We acknowledge that the automated evaluation approach with LLMs used in the
SIGIR LiveRAG 2025 challenge has limitations
\cite{faggioli2022smare,alaofi2025generative,dietz2025llm}. Higher Relevance and
Faithfulness scores do not necessarily mean higher user satisfaction, and
further validation with human annotations or user studies is needed.

It is also worth noting that we have not studied the unintended bias that may
get amplified by the use of LLMs in the different stages of G-RAG, including the
generation of hypothetical answer, query variants, and re-ranking.

\begin{acks}
  We thank the SIGIR 2025 LiveRAG Challenge organizers for the opportunity to participate
  and their support, and the reviewers for their helpful feedback. This research
  was conducted by the ARC Centre of Excellence for Automated Decision-Making
  and Society (ADM+S, \grantnum{ARC}{CE200100005}), and funded fully by the
  Australian Government through the \grantsponsor{ARC}{Australian Research
  Council}{https://www.arc.gov.au/} and was undertaken with the assistance of
  computing resources from RACE (RMIT AWS Cloud Supercomputing).
  This work was conducted on the unceded lands of the  Woi wurrung and Boon
  wurrung language groups of the eastern Kulin Nation. We pay our respect to
  Ancestors and Elders, past and present, and extend that respect to all
  Aboriginal and Torres Strait Islander peoples today and their connections to
  land, sea, sky, and community.
\end{acks}

\clearpage

%\balance

\bibliographystyle{ACM-Reference-Format}
\bibliography{99-references}

\appendix
\label{sec:appendix}

\section{Prompts}
\label{sec:appendix_prompts}

%The main prompts we have used in Section \ref{sec:anova_test} encompass two
%parties: query generation prompts and answer generation prompts. For each
%party, we include three prompts.

\subsection{Query Generation} \label{sec:appendix_prompts_query_generation}

\subsubsection{Na\"ive Prompt}
\label{sec:appendix_prompts_query_generation_naive}

\paragraph{System Prompt:}

\prompt{ Generate a list of \{k\_queries\} search query variants based on the
    user's question, give me one query variant per line. There are no spelling
    mistakes in the original question. Do not include any other text. }

\paragraph{User Prompt:}

\prompt{\{question\}}

\subsubsection{Medium Prompt}
\label{sec:appendix_prompts_query_generation_medium}

Inspired by \citet{min2025promptingalignmentgenerativeframework}, although the
task in this paper is not completely aligned with query variants generation, we
adapted it into a query generation task.
\paragraph{System Prompt:}

\prompt{ You are an expert in query generation, you will be given a question,
    please generate \{k\_queries\} relevant queries based on the question. Make
    sure every query generated can yield new information when I use them to
    search. NEVER repeat similar search queries. }

\paragraph{User Prompt:}

\prompt{Original question: \{question\}}

\subsubsection{Advanced Prompt}
\label{sec:appendix_prompts_query_generation_advanced}

\paragraph{System Prompt:}

\prompt{ Generate \{k\_queries\} diverse search query variations for the given
    question. Follow these guidelines:\\
    1. Each query should focus on different aspects or interpretations of the
    original question\\
    2. Use synonyms and related terms where appropriate\\
    3. Include both broad and specific variations\\
    4. Maintain the core meaning while varying the expression\\
    5. Write each query on a new line\\
    6. Do not include any additional text or formatting\\
    %\\
    The original question is correctly spelled. }

\paragraph{User Prompt:}

\prompt{ Question to analyze: \{question\}\\
    Please generate diverse query variations that capture different aspects of
    this question: }

\subsubsection{Hypothetical Answer Generation}
\label{sec:appendix_prompts_query_generation_hyde}

\paragraph{System Prompt:}

\prompt{Given the question, write a short hypothetical answer that could be true. Be brief and concise.}

\paragraph{User Prompt:}

\prompt{\{question\}}

\subsubsection{Question Decomposition}
\label{sec:appendix_prompts_query_generation_query_decomposition}

\paragraph{System Prompt:}

\prompt{ You are an experienced Google search user, help the user breaking down
    a search question into key components with shorthand entity annotation in
    numbered list style }

\paragraph{User Prompt:}

\prompt{Question: \{question\}}

\subsubsection{Query-to-Question Rephrasing}
\label{sec:appendix_prompts_query_generation_query_rephrasing}

\paragraph{System Prompt:}

\prompt{ You are an experienced Google search user, help the search engine to
    find the results user wanted. Given the main question and its components
    analysis, rephrase into a longer question. What does the user really want? }

\paragraph{User Prompt:}

\prompt{ Question: \{question\}\\
    % \\
    \{components\_str\} }

\subsubsection{Question Classifier}
\label{sec:appendix_prompts_query_generation_is_simple}

\paragraph{System Prompt:}

\prompt{ You are an experienced Google search user, help the user determine if
    the search question is a simple question or a composite question that
    consists of multiple sub-questions. If it's a simple question, you should
    respond: SIMPLE, if it's a composite question, you should respond:
    COMPOSITE. }

\paragraph{User Prompt:}

\prompt{Question: \{question\}}

\subsubsection{Sub-question-to-Query Rephrasing}
\label{sec:appendix_prompts_query_generation_sub_questions}

\paragraph{System Prompt:}

\prompt{ You are an experienced Google search user, help the user to answer the
    question. Given the main question, for each sub-question, create a search
    query, row by row. Your generated query must start with: query: }

\paragraph{User Prompt:}

\prompt{Question: \{question\}}

% --------------------------------------------------------------------------------------------
\subsection{Documents Re-ranking} \label{sec:appendix_reranking}

\subsubsection{Pointwise LLM-based Re-ranker}
\label{sec:appendix_reranking_pointwise}

\paragraph{System Prompt:}

\prompt{ You are a helpful assistant that determines if a document contains
    information that helps answer a given question. Answer only with 'Yes' or
    'No'. }

\paragraph{User Prompt:}

\prompt{ Document: \{doc\_text\}\\
    %\\
    Question: \{question\}\\
    %\\
    Does this document contain information that helps answer this question (only
    answer `Yes' or `No')? }

\subsection{Answer Generation} \label{sec:appendix_answer_gen}

\subsubsection{Na\"ive Prompt} \label{sec:appendix_answer_gen_naive}

\paragraph{System Prompt:}

\prompt{You are a helpful assistant. Answer the question based on the provided documents.}

\paragraph{User Prompt:}

\prompt{ Documents: \{context\}\\
    Question: \{question\}\\
    Answer: }

\subsubsection{Medium Prompt} \label{sec:appendix_answer_gen_medium}

We adapted \citet{CuconasuNoise2024} to support unanswerable questions in the
LiveRAG challenge, e.g., we replaced ``No-RES'' with ``I don't know''. The
reason is that LiveRAG challenge suggests that if the system does not know the
answer, it is better to generate ``I don't know'', instead of the wrong answer.
%,  for the sake of the Relevance evaluation. -- \textit{0 point} stands for
%\textit{No answer is provided in the response (e.g., “I don’t know”)}.

\paragraph{System Prompt:}

\prompt{You are given a question and you MUST respond by EXTRACTING the answer from provided documents. If none of the documents contain the answer, respond with *`I don't know'*.}

\paragraph{User Prompt:}

\prompt{ Documents: \{context\}\\
    Question: \{question\}\\
    Answer: }

\subsubsection{Advanced Prompt}
\label{sec:appendix_answer_gen_advanced}

We adapted \citeauthor{tamber2025benchmarking}'s prompt
\cite{tamber2025benchmarking}, used for their LLM-as-a-judge approach --
FaithJudge.\footnote{\url{https://github.com/vectara/FaithJudge}}
%We used the prompt as described in their GitHub
%repository.\footnote{\url{https://github.com/vectara/FaithJudge}.}

\paragraph{System Prompt:}

\prompt{You must respond based strictly on the information in provided passages. Do not incorporate any external knowledge or infer any details beyond what is given in the passages.}

\paragraph{User Prompt:}

\prompt{ Provide a concise answer to the following question based on the
    information in the provided documents. Documents:\\
    \{context\}\\
    *Question: \{question\}*\\
    Answer: }

\subsection{LLM-based Evaluation}
\label{sec:appendix_evaluators_llm}

We are evaluating the relevance and faithfulness scores of an answer and the
corresponding document rankings separately, in different LLM requests with
different sets of prompts.

\subsubsection{Relevance Score}

\paragraph{System Prompt:}

\prompt{ You are an expert evaluator for Retrieval-Augmented Generation (RAG)
    systems.\\
    Your task is to assess the quality of responses generated by a RAG system
    based on the relevance (correctness) criteria:\\
    Relevance - Measures the correctness and relevance of the answer to the
    question on a four-point scale:\\
    2: The response correctly answers the user question and contains no
    irrelevant content\\
    1: The response provides a useful answer to the user question, but may
    contain irrelevant content that do not harm the usefulness of the answer\\
    0: No answer is provided in the response (e.g., "I don't know")\\
    -1: The response does not answer the question whatsoever\\
    You will be provided with:\\
    - A question\\
    - The response generated by the RAG system\\
    - The retrieved documents used as context\\
    - A gold reference answer (if available)\\
    When a gold reference answer is provided, use it as an additional reference
    point for evaluating the correctness and completeness of the RAG system's
    response. The gold reference represents an ideal answer to the question.\\
    Provide your evaluation in a structured JSON format with the following
    fields:\\
    - evaluation\_notes: Brief explanation of your reasoning for each score\\
    - relevance\_score: The relevance score (-1, 0, 1, or 2)\\
    Be objective and thorough in your assessment. Focus on whether the response
    correctly answers the question. }

\paragraph{User Prompt:}

\prompt{ Please evaluate the following RAG system response:\\
QUESTION:\\
\{question\}\\
RESPONSE:\\
\{answer\}\\
GOLD REFERENCE ANSWER:\\
\{reference\_answer\}\\
RETRIEVED DOCUMENTS:\\
\{documents\}\\
Based on the above, please evaluate the response on relevance (2, 1, 0, or
-1).\\
Provide your evaluation in the following JSON format:\\
\textasciigrave\textasciigrave\textasciigrave json\\
\{\{\\
"evaluation\_notes": "[your reasoning in a single paragraph]",\\
"relevance\_score": [score]\\
\}\}\\
\textasciigrave\textasciigrave\textasciigrave }

\subsubsection{Faithfulness Score}

\paragraph{System Prompt:}

\prompt{ You are an expert evaluator for Retrieval-Augmented Generation (RAG)
    systems.\\
    Your task is to assess the quality of responses generated by a RAG system
    based on the faithfulness (support) criteria:\\
    Assess whether the response is grounded in the retrieved passages on a
    three-point scale:\\
    1: Full support, all answer parts are grounded\\
    0: Partial support, not all answer parts are grounded\\
    -1: No support, all answer parts are not grounded\\
    You will be provided with:\\
    - A question\\
    - The response generated by the RAG system\\
    - The retrieved documents used as context\\
    Provide your evaluation in a structured JSON format with the following
    fields:\\
    - evaluation\_notes: Brief explanation of your reasoning for each score\\
    - faithfulness\_score: The faithfulness score (-1, 0, or 1)\\
    Be objective and thorough in your assessment. Focus on whether the response
    correctly answers the question and is supported by the retrieved documents.
    }

\paragraph{User Prompt:}

\prompt{ Please evaluate the following RAG system response:\\
QUESTION:\\
\{question\}\\
RESPONSE:\\
\{answer\}\\
GOLD REFERENCE ANSWER:\\
\{reference\_answer\}\\
RETRIEVED DOCUMENTS:\\
\{documents\}\\
Based on the above, please evaluate the response on faithfulness (1, 0, or
-1).\\
Provide your evaluation in the following JSON format:\\
\textasciigrave\textasciigrave\textasciigrave json\\
\{\{\\
"evaluation\_notes": "[your reasoning in a single paragraph]",\\
"faithfulness\_score": [score]\\
\}\}\\
\textasciigrave\textasciigrave\textasciigrave }

\section{CRediT Author Statements}

% The CRediT taxonomy
% - Conceptualization
% - Data curation
% - Formal analysis
% - Funding acquisition
% - Investigation
% - Methodology
% - Project administration
% - Resources
% - Software
% - Supervision
% - Validation
% - Visualization
% - Writing -- original draft
% - Writing -- review \& editing Explanation and examples for each category are
%   available here: https://credit.niso.org/ 

% Enter the roles for each author as a comma-separated list with exact spelling.
% Order can be arbitrary.
\credit{KR}{Conceptualization, Methodology, Software, Data curation, Formal analysis, Investigation, Writing -- original draft, Writing -- review \& editing}
\credit{SS}{Conceptualization, Data curation, Formal analysis, Investigation, Methodology, Software, Validation, Visualization, Writing -- original draft, Writing -- review \& editing}
\credit{KNDA}{Methodology, Investigation, Software, Writing -- review \& editing}
\credit{DS}{Conceptualization, Formal analysis, Investigation, Methodology, Supervision, Visualization, Writing -- original draft, Writing -- review \& editing}
\credit{OZ}{Conceptualization, Formal analysis, Funding acquisition, Resources, Investigation, Methodology, Supervision, Writing -- review \& editing}

\insertcredits
%\\

\noindent \insertcreditsstatement.

\end{document}